\documentclass[twocolumn,amsmath,amssymb,aps,prl]{revtex4}
\usepackage{graphicx}% Include figure files

\gdef\ffrac#1#2{\textstyle{#1\over#2}\displaystyle}

\begin{document}
%\begin{fmffile}{pics}
\pagestyle{empty}
\preprint{NSF-KITP-14-021}

\title{Thermalization and Revivals after a Quantum Quench in Conformal Field Theory}

\author{John Cardy}
\affiliation{Rudolf Peierls Centre for Theoretical Physics, Oxford
University, 1 Keble Road, Oxford, OX1 3NP, United Kingdom}
\affiliation{All Souls College, Oxford, United Kingdom}
\affiliation{Kavli Institute for Theoretical Physics, Santa Barbara}

%\date{March 7 2011}% It is always \today, today,
\date{March 11, 2014}

\begin{abstract}
We consider a quantum quench in a finite system of length $L$ described by a 1+1-dimensional CFT, of central charge $c$, from a state with finite energy density corresponding to an inverse temperature $\beta\ll L$. For times $t$ such that $\ell/2<t<(L-\ell)/2$ the reduced density matrix of a subsystem of length $\ell$ is exponentially close to a thermal density matrix.
We compute exactly the overlap $\cal F$ of the state at time $t$ with the initial state and show that in general it is exponentially suppressed at large $L/\beta$.  However, for minimal models with $c<1$ (more generally, rational CFTs),
at times which are integer multiples of $L/2$ (for periodic boundary conditions, $L$ for open boundary conditions)
there are (in general, partial) revivals at which $\cal F$ is $O(1)$, leading to an eventual complete revival with ${\cal F}=1$. There is also interesting structure at all rational values of $t/L$, related to properties of the CFT under modular transformations.  At early times $t\!\ll\!(L\beta)^{1/2}$ there is a universal decay ${\cal F}\sim\exp\big(\!-\!(\pi c/3)Lt^2/\beta(\beta^2+4t^2)\big)$. The effect of an irrelevant non-integrable perturbation of the CFT is to progressively broaden each revival at $t=nL/2$ by an amount $O(n^{1/2})$. 

\end{abstract}

%\pacs{05.30.Rt, 03.67.Mn}% PACS, the Physics and Astronomy
%\keywords{Entanglement entropy, Quantum critical behavior}%Use showkeys class option if keyword
\maketitle

The subject of quantum quenches, the time evolution of an extended system, described by a hamiltonian $H$, from a pure state $|\psi_0\rangle$ which is not an eigenstate (usually the ground state of some other hamiltonian $H_0$), has been of great interest in recent years, both for theoretical reasons and the fact that such coherent evolution may be experimentally realised in ultracold atoms. Important theoretical questions are whether, and in what sense, such systems reach a stationary state, and to what extent this can be described by a thermal density matrix. These are difficult to address except in theories which are in some way exactly solvable \cite{CC,quench}, or in the AdS/CFT correspondence, when thermalization has been associated with the formation of a black hole in the bulk \cite{BH}. 

In \cite{CC} the problem was studied for the case when $H$ corresponds to a 1+1-dimensional conformal field theory (CFT) and $|\psi_0\rangle$ is a particular kind of initial state with short-range correlations and entanglement. It was found that correlation functions of local observables within a subsystem of length $\ell$ become stationary after a time $\approx\ell/2$ (in units where the speed of propagation is unity), after which they are described by a thermal ensemble at a temperature corresponding to the conserved energy density. At the same time the entanglement entropy of the subsystem with its complement becomes equal to the Gibbs entropy at the same temperature. These results may be explained within a simple physical picture of pairs of left- and right-moving quasiparticles, initially entangled over a length scale $\sim\beta$, being emitted at $t=0$ and thereafter moving semi-classically. This general picture has been confirmed in other integrable lattice models, although in these cases the stationary state is a generalized Gibbs ensemble (GGE) rather than a purely thermal state \cite{GGE}. 

These considerations have largely been made for the thermodynamic limit, when the total length $L$ of the system is first taken to infinity. However, for a finite system, the quasiparticle picture also implies quantum recurrence. In a periodic system an oppositely moving pair of particles will meet again at times which are integer multiples of $L/2$, and this, in the absence of accidental destructive interference, should lead to a revival of the initial state. In open systems with reflecting boundaries, such revivals should occur at multiples of $L$. In some integrable quantum spin chains such revivals in the expectation values of local observables have indeed been observed \cite{Igloi} . 

In this Letter we describe the extension of the methods developed in \cite{CC} to the case of finite systems. With the same assumptions about the form of the initial state, we first make precise the statement of thermalization, by computing the overlap between the reduced density matrix of a subsystem of length $\ell$ at time $t$ and that of a thermal mixed state. We find that for times $t>\ell/2$ this is exponentially close to unity as $(t-\ell/2)/\beta\to\infty$. This holds up to a time $\sim(L-\ell)/2$ when the subsystem recoheres. To understand this further, we compute exactly the overlap, or fidelity ${\cal F}(t)=|\langle\psi_0|e^{-iHt}|\psi_0\rangle|$ of the quantum state at time $t$ with the initial state, by relating this quantity to the partition function of the CFT on an annulus (or rectangle for open boundary conditions) continued to complex values of its modulus or aspect ratio. Since much is known about these partition functions (in some cases completely) we are able to obtain a number of analytic results. We note in passing that in recent papers \cite{zeroes} a similar quantity has been studied as a function of complex $t$ for various spin chains, and its singularities interpreted as `phase transitions' at finite $t$. For the case of a CFT studied here, the singularities we find occur close to every rational value of $t/L$ and are simply related to full or partial revivals of the initial state.

\noindent\em Formulation of the problem. \em 
In principle, the initial state $|\psi_0\rangle$ should be the ground state of a perturbed hamiltonian $H+\lambda\int\!\Phi dx$, where $\lambda$ is a relevant coupling to a local operator $\Phi$ which gaps the system, leading to a finite correlation length  which we assume is always $\ll L$. In practice, this is too difficult, and instead we assume \cite{CC} that $|\psi_0\rangle$ is close in the renormalization group sense to some conformal boundary state $|B\rangle$. However since such states are scale invariant (and not even normalizable), in order to introduce a finite correlation length we take instead $|\psi_0\rangle\propto e^{-(\beta/4)H}|B\rangle$.
This somewhat arbitrary choice was motivated on phenomenological grounds in  \cite{CC}, but a better argument is to point out that $H\propto\int T_{tt}dx$, where $T_{tt}$ is the local stress tensor, is (often the most leading) irrelevant operator which acts on the boundary state. $\beta$, which initially appears here only as a coupling constant, is in fact chosen so that the mean energy $\langle\psi_0|H|\psi_0\rangle=\pi cL/24(\beta/2)^2$ is the same as that in a thermal state $\pi cL/6\beta^2$ \cite{BCN}. The effect of modifying the initial state by adding other irrelevant operators may be argued to lead to the stationary state being described by a GGE rather than a purely thermal one \cite{JCGGE}.

\noindent\em Thermalization of a subsystem. \em
Consider a bipartition of the Hilbert space ${\cal H}_A\otimes{\cal H}_B$ into the degrees of freedom in the interval $A:|x|\!<\!\ell/2$ and the remainder $\ell/2\!<\!|x|\!<\!L/2$. The reduced density matrix of $A$ at imaginary time $\tau$ is then
$\rho_A(\beta,\tau)\propto{\rm Tr}_{{\cal H}_B}(e^{-\tau H}e^{-\frac14\beta H}|B\rangle\langle B|e^{-\frac14\beta H}e^{\tau H})$. Following \cite{cc1}, this may be thought of as the partition function of the CFT on an annulus ${\cal A}:-\frac14\beta<\tau<\frac14\beta$ times a circle of circumference $L$, cut open along $(\tau,|x|<\frac12\ell)$. 
Similarly, the thermal density matrix $\tilde\rho_A(\beta)\propto{\rm Tr}_{{\cal H}_B}\,e^{-\beta H}$ is given by the partition function on a torus ${\cal T}:-\frac12\beta<\tau<\frac12\beta$ times the same circle, cut open in the same manner. An estimate of their closeness is
$$
I(\tau)\equiv\frac{{\rm Tr}_{{\cal H}_A}\big(\rho_A(\beta,\tau)\tilde\rho_A(\beta)\big)}{\big({\rm Tr}_{{\cal H}_A}(\rho_A(\beta,\tau)^2)\,{\rm Tr}_{{\cal H}_A}(\tilde\rho_A(\beta)^2)\big)^{1/2}}\leq1\,,
$$
with equality only when the two density matrices are identical. The numerator is given by the partition function
$Z_{{\cal A}\oplus{\cal T}}$ on the surface formed by sewing together $\cal A$ and $\cal T$ along the cut, and similarly the two factors in the denominator are given by $Z_{{\cal A}\oplus{\cal A}}$ and $Z_{{\cal T}\oplus{\cal T}}$ respectively.
In general these are difficult to evaluate. However if we set $w\equiv x+i\tau$ and consider the conformal mapping $w\to z=ie^{2\pi w/\beta}$, then as $L\to\infty$ $\cal A$ is mapped into the upper half $z$-plane and $\cal T$ into the whole plane.
On continuing $\tau\to it$, the ends of the interval $(\tau,|x|<\frac12\ell)$ are mapped to $z=ie^{(2\pi/\beta)(\pm\frac12\ell-t)}$. 
Similarly the image points in the real axis lie at $\bar z=-ie^{(2\pi/\beta)(\pm\frac12\ell+t)}$.
As discussed in \cite{CC} for the similar problem of the 2-point function of a local observable, there are two regimes. If $\frac12\ell-t\gg\beta$, each point is exponentially close its image compared to its distance from the other point. If $t-\frac12\ell\gg\beta$, they are exponentially close together compared to their distance to their images.
In the latter case we may ignore the boundary, which is equivalent to replacing ${\cal A}$ by a cylinder $\cal C$. In this approximation $I=1$. The corrections to this come from the existence of the boundary, and may be estimated using the short interval expansion developed in \cite{CCT}, giving
$$
1-I\sim e^{-4\pi\Delta_{\rm min}(t-\ell/2)/\beta}\,,
$$
where $\Delta_{\rm min}$ is the smallest dimension among those operators which have a non-zero expectation value in the initial state.

The above assumes $L\gg t,\ell$. However, if there is (partial) revival at $t=L/2$ (as we argue below) the same argument working backwards shows that thermalization should begin to fail once $t>(L-\ell)/2$.

\noindent\em Return amplitude. \em
With the above choice for $|\psi_0\rangle$, the return amplitude is
\begin{equation}\label{eq1}
\!{\cal F}\!=\!\left|\frac{\langle B|e^{-\frac14\beta H}e^{-itH}e^{-\frac14\beta H}|B\rangle}
{\langle B|e^{-\frac14\beta H}e^{-\frac14\beta H}|B\rangle}\right|
\!=\!\left|\frac{Z_{\cal A}(\ffrac12\beta+it,L)}{Z_{\cal A}(\ffrac12\beta,L)}\right|\!\!\!\!\!\!
\end{equation}
where $Z_{\cal A}(W,L)$ is the partition function of the CFT on an annulus of width $W$ and circumference $L$, with conformal boundary conditions corresponding to $B$ on both edges. A great deal is known about the form of $Z_{\cal A}$ for a 
CFT \cite{JC89}: 
\begin{equation}\label{eq2}
Z_{\cal A}(W,L)=\sum_\Delta |B_\Delta|^2\chi_\Delta(q)=\sum_{\widetilde\Delta}n_{BB}^{\widetilde\Delta}
\chi_{\widetilde\Delta}(\tilde q)\,,
\end{equation}
where $q\equiv e^{2\pi i\tau}=e^{-4\pi W/L}$, $\tilde q=e^{-2\pi i/\tau}=e^{-\pi L/W}$, and $\Delta,\widetilde\Delta$ label the highest weights of Virasoro representations which propagate across and around the annulus respectively. $\chi_\Delta(q)=q^{-c/24+\Delta}\sum_{N=0}^\infty d_Nq^N$ are the characters of these representations, where $d_N$ is their degeneracy at level $N$. The coefficients $B_\Delta$ are the overlaps between the physical states $B$ and the Ishibashi states \cite{Ishi}. 
The non-negative integers $n_{BB}^{\widetilde\Delta}$, which for a rational CFT are given by the fusion rules, give the number of states with highest weight $\widetilde\Delta$ allowed to propagate around the annulus with the given boundary conditions.  We assume $n_{BB}^0=1$. For minimal CFTs with $c<1$ there is a finite number of allowed values of $\Delta$ and $\widetilde\Delta$ given by the Kac formula. For more general rational CFTs the number of different values (mod $\bf Z$) is still finite, but for a general CFT with $c>1$ it is infinite, the mean density growing exponentially with $\sqrt\Delta$ \cite{JC86}.  

The main property of the characters which we need is that they are holomorphic in the upper half $\tau$-plane, and that they transform linearly under a representation of the modular group SL$(2,{\bf Z})$, generated by $S:\tau\to-1/\tau$ and 
$T:\tau\to\tau+1$. The first property ensures that the continuation to $\tau=(-2t+i\beta)/L$ implied in (\ref{eq1}) makes sense, and the second will allow us to relate the values of ${\cal F}(t)$ at different times to those back in the principal domain where $\tau\to i\infty$ and the series are rapidly convergent.

\noindent\em Universal short time behaviour. \em 
Note that $\tilde q=\exp\big(\!-2\pi L(\beta\!-\!2it)/(\beta^2+4t^2)\big)$.
For $t^2\ll L\beta$, $|\tilde q|\ll1$, and so the sum on the rhs of (\ref{eq2}) is dominated by its first term ${\tilde q}^{-c/24}$. 
After normalising by the denominator in (\ref{eq1}) this gives the first main result
\begin{equation}\label{eq3}
\!{\cal F}(t)\sim\exp\!\big(\!-\!(\pi c/3)Lt^2/\beta(\beta^2\!+\!4t^2)\big)\left(1\!+\!O(|\tilde q|^{\alpha})\right)\,,
\end{equation}
which shows a decay, initially faster than exponential, to a plateau value which is however exponentially small in $L/\beta$. The power $\alpha$ in the correction term is the smallest non-zero value of $\widetilde\Delta$ such that $n_{BB}^{\widetilde\Delta}\geq1$, or 2. We stress that this result should hold for \em any \em CFT.

\noindent\em Revival. \em
$t=2n/L$ corresponds to $\tau\approx-n$, and we may then relate the value of $Z_{\cal A}$ at this point to that near $\tau=0$, and then as $\tau\to i\infty$ using the transformation properties of the characters. This gives, in the limit $L/\beta\to\infty$, 
$$
{\cal F}(nL/2)=\sum_\Delta|B_\Delta|^2\left({\bf T}^n{\bf S}\right)_{\Delta,0}=\sum_{\widetilde\Delta}n_{BB}^{\widetilde\Delta}
\left({\bf S}{\bf T}^n{\bf S}\right)_{\widetilde\Delta,0}\,,
$$
where $\bf S$ and $\bf T$ are the corresponding matrices according to which the characters transform. 
It follows that as long these are finite dimensional (as for the the minimal models or more generally a rational CFT), the value of ${\cal F}(t)$ at $t=nL/2$ is therefore finite (although, as we shall see below, it may accidentally vanish). At times within $(L\beta)^{1/2}$ of this there is a similar decay to that in (\ref{eq3}) with $t$ replaced by $|t-nL/2|$. If $M$ is the lowest common denominator of all the $\{\Delta\}$, then, since all the energy gaps of $H$ (of even parity) are quantized in units of $4\pi/LM$, there must always be complete revival (${\cal F}=1$) at multiples 
of $t=ML/2$. For the minimal models, the Kac formula implies that in $M\sim24/(1-c)$ and therefore in general the time for a complete revival diverges as $c\to1-$. We also find (numerically) that in the same limit the return amplitude at any fixed revival time goes to zero exponentially fast. 
A similar result should hold for other sequences of rational CFTs with a maximal value of $c$. 

\begin{figure}
\includegraphics[width=0.4\textwidth]{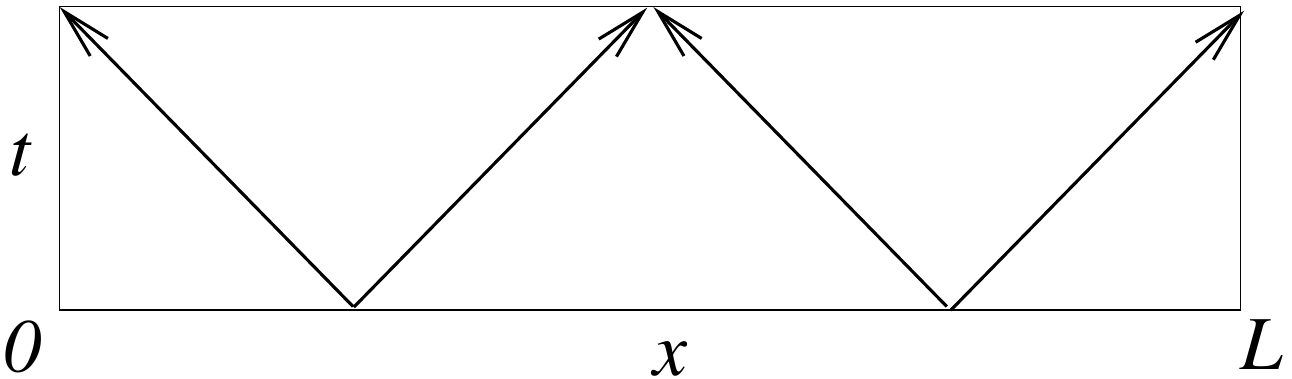}
\caption{\label{fig1}Quasiparticle configuration leading to the feature in the return amplitude at $t=L/4$ for periodic boundary conditions. The pairs emitted a distance $L/2$ apart must be correlated, leading to an exponential suppression.}
\end{figure}

\noindent\em Structure at rational values of $t/L$. \em
Although finite values of ${\cal F}$  occur only at integer values of $2t/L$, in fact there is interesting universal structure near every rational value. This is because the characters are singular at $\tau=0$, and the modular group maps this to every rational point $\tau=n/m$ on the real line.
This is mapped to $\tau=0$ by applying $ST^{n_1}ST^{n_2}\ldots$, where $(n_1,n_2,\ldots)$ are the integers appearing in the continued fraction expansion of $n/m$. However the nearby point $\tau=n/m+i\beta/L$ in mapped to $\tau\approx im^2(\beta/L)$ and so we find, after normalising with the denominator of (\ref{eq1}),
\begin{equation}\label{eq4}
{\cal F}(nL/2m)\propto (e^{-2\pi L/\beta})^{(c/24)(1-1/m^2)}\,.
\end{equation}
Once again, at nearby values of $t$, this is modified in a similar manner to (\ref{eq3}). A more careful analysis also shows that the correction terms may be neglected only for $m\ll(L/\beta)^{1/2}$, so that for a fixed $\beta/L$ the structure near only a finite number of rational values will be evident. This result however shows that if we define a `large deviation function'
$-\lim_{L/\beta\to\infty}(\beta/2\pi L)\log{\cal F}(t)$, it is a sum of delta functions of strength $\propto1/m^2$ at each rational value $n/m$ of $2t/L$, on top of the uniform plateau value $c/24$.
This structure may be understood in the quasiparticle picture as being due to the simultaneous emission at $t=0$ of entangled pairs of particles separated by distances which are integer divisors of $L$. An example is illustrated in Fig.~\ref{fig1}. 

\noindent\em Example: Ising CFT. \em
Many of these features are present in the simplest minimal CFT, corresponding to the scaling limit of the Ising model with $c=\frac12$. There are three distinct conformal boundary states, corresponding to the scaling limits of free and fixed boundary conditions on the Ising spins. In the last two cases \cite{JC89}, corresponding to a quench in the transverse field Ising model to the critical point from the ground state in a large longitudinal field, or from the ordered phase,
$$
Z_{\cal A}^{\rm fixed}=\ffrac12\chi_0(q)+\ffrac12\chi_{1/2}(q)+\ffrac1{\sqrt2}\chi_{1/16}(q)=\chi_0(\tilde q)\,.
$$
At the recurrence times $t=nL/2$, we find by applying ${\bf T}^{n}$ and then $\bf S$
\begin{eqnarray*}
Z_{\cal A}^{\rm fixed}&=&\ffrac12\chi_0(q')+\ffrac12e^{i\pi n}\chi_{1/2}(q')+\ffrac1{\sqrt2}e^{i\pi n/8}\chi_{1/16}(q)\\
&\sim&\left[\ffrac14(1+e^{i\pi n})+\ffrac12e^{i\pi n/8}\right]\chi_0(\tilde{q'})\,,
\end{eqnarray*}
where $q'=e^{-2\pi\beta/L}$, $\tilde{q'}=e^{-2\pi L/\beta}$, and we have retained only the dominant term in the second step.
For $n$ odd this gives ${\cal F}(nL/2)=\frac12$, while for $n$ even we get $|\cos(\pi n/16)|$. There is complete revival at $t=8L$, while at $t=4L$ the coefficient vanishes, leaving a much smaller term $O((e^{-2\pi L/\beta})^{1/16})$. 

\begin{figure}
\includegraphics[width=0.45\textwidth]{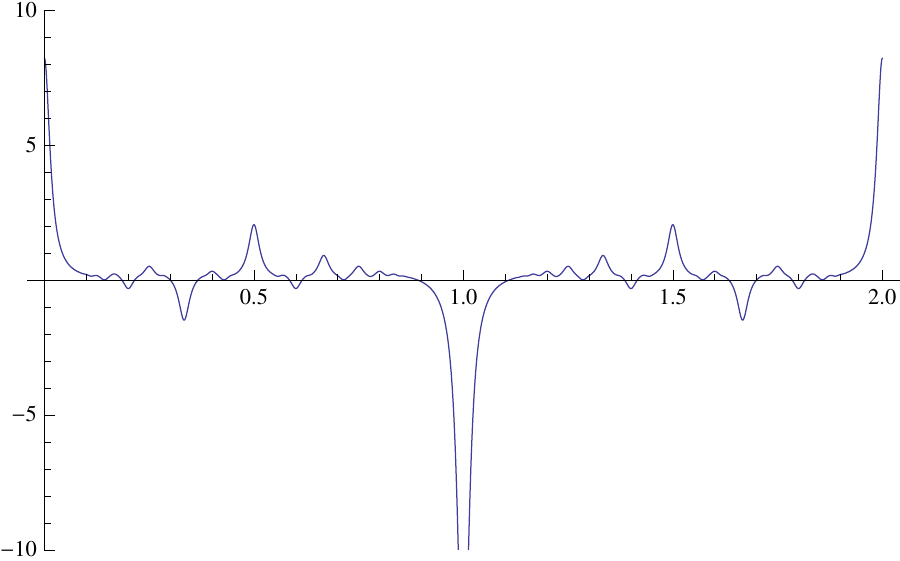}
\caption{\label{fig2}Log of the return amplitude for the Ising CFT starting from a disordered state for $0<2t/L<2$, with $\pi\beta/L=0.1$. The vertical axis has been shifted so as to expose the mean plateau behaviour. This shows the initial gaussian decay and revival at $t=L$. The negative peak at $t=L/2$ is due to destructive interference between two kinds of quasiparticles. Smaller gaussian peaks are seen at rational values with small denominators.
The positive peaks are mapped by the modular group to the initial peak, and the negative ones to the feature at $2t/L=1$.}
\end{figure}

\begin{figure}
\includegraphics[width=0.45\textwidth]{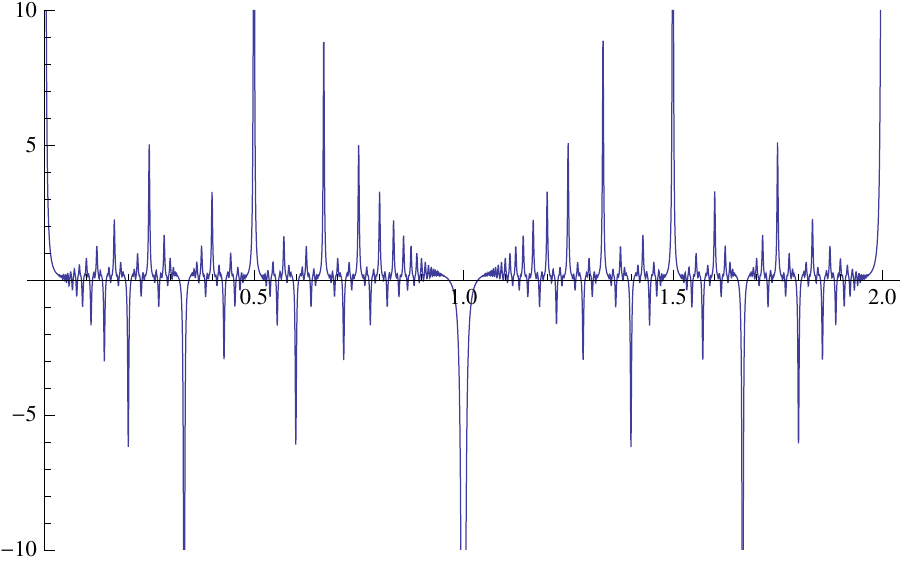}
\caption{\label{fig3} Same as above with $\pi\beta/L=0.01$. Now there is structure at more rational values, and we see the predicted $1/m^2$ dependence of the heights of nearby peaks with denominators $m$.}
\end{figure}

On the other hand, for free boundary conditions \cite{JC89}, corresponding to a quench from the disordered phase in zero longitudinal field,
$$
Z_{\cal A}^{\rm free}=\chi_0(q)+\chi_{1/2}(q)=\chi_0(\tilde q)+\chi_{1/2}(\tilde q)\,.
$$
At $t=nL/2$, we get $\chi_0(q')+(-1)^n\chi_{1/2}(q')$, so for $n$ even there is complete revival, but, for odd $n$,
$\chi_0-\chi_{1/2}=\sqrt2\chi_{1/16}(\tilde{q'})$, so again the revival is suppressed. The above expression may also be written as 
$Z_{\cal A}^{\rm free}=q^{-1/48}\prod_{k=0}^\infty(1+q^{k+1/2})$, which explicitly shows the structure near rational values of $2t/L$. This is illustrated in Figs.~(\ref{fig2}, \ref{fig3}).

\noindent\em Open boundary conditions. \em
Suppose now that the system is open with conformal boundary conditions $B'$ at $x=\pm L/2$. (We may also introduce an extrapolation length $\ell_0$ in order to smooth out this condition, but this only has the effect of changing $L$ to $L+2\ell_0$ and we shall ignore it.) Then $Z_{\cal A}$ in (\ref{eq1}) is replaced by $Z_{BB'}$, the partition function for a $W\times L$ rectangle. In the special case when $B=B'$, the CFT partition function is known exactly \cite{Kleban1}:
$$
Z_{BB}(W,L)=L^{c/4}\eta(q)^{-c/2}=W^{c/4}\eta(\tilde q)^{-c/2}\,,
$$
where now $q=e^{-2\pi W/L}$, $\tilde q=e^{-2\pi L/W}$, and $\eta(q)=q^{1/24}\prod_{n=1}^\infty(1-q^n)$. Setting $W=\frac12\beta+it$ we see that ${\cal F}(t)$ is now recurs with period $L$. This exact revival may be traced to the fact that, although the spatial boundary conditions may allow other states corresponding to Virasoro representations with $\Delta\not=0$, the initial condition selects only those which are descendants of the identity. An example for the Ising CFT would be to consider a quench from the disordered phase in a system with free boundary conditions on the Ising spins at $x=\pm\frac12L$. However, the result holds for any CFT, whether it is rational or not. The same property, of exact revival at multiples of $\pi/L$, will also occur irrespective of the initial state if $B'$ is such that $n_{B'B'}^\Delta=\delta_{\Delta,0}$. Such boundary conditions are known to exist for all the minimal models, for example fixed boundary spins in the Ising model. The modular properties of $\eta(q)$ imply that there is structure near all rational multiples of $t/L$, similar to the case of periodic boundary conditions.

\noindent\em One-point functions. \em
Analytic results for the 1-point function of a local operator $\Phi$ in a finite annulus or rectangle are available in only a few 
cases \cite{Kleban2}. The simplest is that discussed above, a $W\times L$ rectangle with the same conformal boundary condition on each edge. Since this geometry is conformally equivalent to the upper half plane, where the 1-point function decays as a power $\Delta_\Phi$ of the distance from the real axis, it follows that in the rectangle $\langle\Phi(\tau,x)\rangle$ is a completely universal function of 
the coordinates and $(W,L)$, for any CFT. The simplest and most useful form is then found by taking $\Phi$ to be the exponential of a massless scalar field, for which the method of images may be used. The result, after continuing to real times, and taking $x$ to be at the midpoint for simplicity, is
$$
\langle\Phi(t,0)\rangle\propto
%\left(\frac{2\pi}\beta\right)^\Delta
\prod_{m=-\infty}^\infty\left[\frac{\cosh\big((2\pi/\beta)(t-(m+\frac12)L)\big)}{\cosh\big((2\pi/\beta)(t-mL)\big)}\right]^{\Delta_\Phi}\,.
$$
This shows an exponential decrease, as for the infinite system, until $t\approx \frac12L$, followed by a symmetrical recovery to
the initial value at $t=L$. Note that there is no signal of the fine structure which occurs in the overlap at rational $t/L$, but the quasiparticle picture suggests that this should show up in the higher-point functions.

\noindent\em Non-integrable perturbations. \em
The simple picture of partial and exact revivals at multiples of $L/2$ (in a periodic system) in a pure CFT is clearly a consequence of the integrable structure imposed by the Virasoro algebra. In any realistic critical system, $H$ will contain irrelevant terms which in general spoil the integrability. In general their effect is very difficult to quantify. However some progress is possible for an irrelevant perturbation of the form $\delta H=\lambda\int T\overline Tdx$, which, for many systems, is the most important scalar irrelevant operator. In a periodic system of size $L$, in first-order perturbation theory it causes a shift $\sim(\lambda/L^3)
(-c/24+\Delta)^2$ in a level whose unperturbed energy is $\sim(4\pi/L)(-c/24+\Delta)$, but, to this order, the degeneracies remain. Thus the characters $\chi_\Delta(q)$ appearing in the first expression in (\ref{eq2}) are replaced by
$$
\chi_\Delta (q)\to\sum_Nd_Nq^{-\frac c{24}+\Delta+N+(\lambda/L^2)(-\frac c{24}+\Delta+N)^2}\,.
$$
Writing the quadratic term as a gaussian integral $\propto\int d\xi\,q^{\xi^2+i((\lambda^{1/2})/2L)\xi(-\frac c{24}+\Delta+N)}$,
we see that we
may take the expressions for ${\cal F}$ evaluated within the pure CFT at times $t(1+O(\lambda^{1/2}\xi/L))$ and integrate them against $q^{\xi^2}\sim e^{-2\pi(\beta+2it)\xi^2/L}$. This will lead to an $O(n^{1/2}\lambda^{1/2})$ broadening of the revival peak at $t=nL/2$. (There is also a $\xi$-dependent shift in $\beta$, which makes the peaks asymmetrical.)
At $O(\lambda^2)$ the degeneracies are split, leading to a new time scale $O(L^5/\lambda^2)$ beyond which we would expect to see complete decoherence.

\noindent\em Discussion. \em
1+1-dimensional CFTs in a finite system have spectral gaps which (at zero momentum, in periodic systems) are integer multiples of $4\pi/L$, which naturally leads to revivals at times which are multiples of $L/2$. However the spectrum is purely of this form only if the initial state is of the Ishibashi form, and in general these are unphysical. For minimal models (more generally, rational CFTs) only finite combinations of these are needed, leading to partial revivals and then full revival at some multiple $M$ of $L/2$. For irrational CFTs with large $c$, on the other hand, an infinite number of Ishibashi states are needed to form the physical states. In this case it is unlikely that a complete revival is possible, but this leaves open the question of whether finite partial revivals may occur. The behaviour of the minimal models as $c\to1-$ suggests that this is not the case. For CFTs with a weakly coupled AdS/CFT dual this is consistent with the idea that a black hole will eventually form in the bulk after possibly some oscillations.

\noindent\em Acknowledgements. \em
The author thanks the following for questions and comments:
P.~Calabrese,  F.~Essler, T.~Hartman, V.~Keranen, P.~Kleban, D.~Marolf, R.~Myers, A.~O'Bannon, E.~Rabinovici and A.~Starinets. 
This research was carried out while visiting the Kavli Institute for Theoretical Physics, Santa Barbara, and was supported by the NSF under Grant No.~NSF PHY11-25915 and by the Simons Foundation.


\begin{thebibliography}{99}
%
\bibitem{CC} P.~Calabrese and J.~Cardy, 
Phys. Rev. Lett. {\bf 96}, 136801 (2006); J. Stat. Mech. 0706:P06008 (2007).
%
\bibitem{quench} Earlier papers were:
E. Barouch, B. McCoy, and M. Dresden, Phys. Rev. A {\bf 2}, 1075 (1970);
Phys. Rev. A {\bf 3}, 786 (1971); Phys. Rev. A {\bf 3}, 2137 (1971);
F. Igloi and H. Rieger, Phys. Rev. Lett. {\bf 85}, 3233 (2000);
K. Sengupta, S. Powell, and S. Sachdev, Phys. Rev. A {\bf 69}, 053616 (2004).
%
\bibitem{BH} U.H. Danielsson, E. Keski-Vakkuri, and M. Kruczenski,
JHEP 0002:039 (2000);
S.~Bhattacharyya and S.~Minwalla, JHEP {\bf 0909}, 034 (2009);
R.~A.~Janik and R.~B.~Peschanski,
Phys. Rev. D {\bf 74}, 046007 (2006);
H.~Ebrahim and M.~Headrick, arXiv:1010.5443 [hep-th];
J.~Abajo-Arrastia, J.~Aparicio and E.~Lopez,
JHEP {\bf 1011}, 149 (2010);
V.~Balasubramanian {\it et al.},
Phys. Rev. Lett. {\bf 106}, 191601 (2011);
A.~Buchel, R.C.~Myers, A. ~van Niekerk,
Phys. Rev. Lett. {\bf 111}, 201602 (2013).
%
\bibitem{GGE} M. Rigol, V. Dunjko, V. Yurovsky, and M. Olshanii, Phys.
Rev. Lett. {\bf 98}, 050405 (2007); 
A. C. Cassidy, C. W. Clark and M. Rigol, Phys. Rev. Lett. {\bf 106}, 140405 (2011);
P. Calabrese, F.H.L. Essler, and M. Fagotti, J. Stat. Mech. P07022 (2012);
M.~Fagotti and F. H. L.~Essler; Phys. Rev. B {\bf 87}, 245107 (2013).
%
\bibitem{Igloi} H.~Rieger and F.~Igl\'oi Phys. Rev. Lett. {\bf 106}, 035701 (2011);
Phys. Rev. B {\bf 84}, 165117 (2011);
J.~H\"app\"ol\"a, G.B.~Hal\'asz, A.~Hamma,
Phys. Rev. A {\bf 85}, 032114 (2012).
%
\bibitem{zeroes} M.~Heyl, A.~Polkovnikov and S.~Kehrein,
Phys. Rev. Lett. {\bf 110}, 135704 (2013);
C.~Karrasch and D.~Schuricht, Phys. Rev. B {\bf 87},195104 (2013);
M.~Fagotti, arXiv:1308.0277;
B.~Pozsgay, arXiv:1308.3087;
F.~Andraschko and J.~Sirker, arXiv:1312.4165;
S.~Vajna and B.~D\'ora, arXiv:1401.2865.
%
\bibitem{BCN} H. W. J.~Bl\"ote, J.~Cardy and M. P.~Nightingale, Phys. Rev. Lett. {\bf 56}, 742 (1986);
I.~Affleck, Phys. Rev. Lett. {\bf 56}, 746 (1986 ).
%
\bibitem{JCGGE} J.~Cardy, in preparation.
%
\bibitem{cc1} P.~Calabrese and J.~Cardy,  J. Stat. Mech. 0406:P06002 (2004).
%
\bibitem{CCT} M.~Headrick, Phys. Rev. D {\bf 82},126010 (2010);
P.~Calabrese, J.~Cardy and E.~Tonni, J. Stat. Mech., P11001 (2009); J. Stat. Mech.,  P01021 (2011).
%
\bibitem{JC89} J.~Cardy, Nucl. Phys. B {\bf 324}, 581 (1989).
%
\bibitem{Ishi} N.~Ishibashi, Mod. Phys. Lett. A {\bf 4}, 251 (1989).
%
\bibitem{JC86} J.~Cardy, Nucl. Phys. B {\bf 270}, 186 (1986).
%
\bibitem{Kleban1} P.~Kleban and I.~Vassileva, J. Phys. A {\bf 24}, 3407 (1991).
%
\bibitem{Kleban2} J. J. H.~Simmons and P.~Kleban, 
 J. Phys. A {\bf 44}, 315403 (2011).
%



\end{thebibliography}
\end{document}